\documentstyle[aps,twocolumn,floats,prl]{revtex} 
\begin{document} 
\draft 
\title{
Asymptotics of Universal Probability of Neighboring Level Spacings\protect\\
at the Anderson Transition
}
\author{Isa Kh. Zharekeshev\cite{byline} and Bernhard Kramer\\}
\address{
1. Institut f\"ur Theoretische Physik, Universit\"at Hamburg,
Jungiusstrasse 9, D-20355 Hamburg, Germany\smallskip\\
\rm{(June 23, 1997; accepted for publication
in Physical Review Letters)}\bigskip\\
\parbox{14.2cm}{\rm
The nearest-neighbor level spacing distribution 
is numerically investigated by directly diagonalizing disordered 
Anderson Hamiltonians for systems of sizes up to 100$\times$100$\times$100 
lattice sites. 
The scaling behavior of the level statistics
is examined for large spacings near the delocalization-localization
transition  and the correlation length exponent is found.
By using high-precision calculations we conjecture a new interpolation
of the critical cumulative probability, which has size-independent
asymptotic form 
$\ln I(s) \propto -s^{\alpha}$
with  
$\alpha = 1.0\pm 0.1$. 
}}

\maketitle
\narrowtext

The statistical fluctuations
in energy spectra of disordered quantum systems 
attract at present much attention \cite
{Shklovskii93,Hofstetter94,Kravtsov94,Aronov94,Evangelou94}. 
It is known that by increasing the fluctuations
of a random potential 
the one-electron states undergo
a localization transition, which is the origin of the
Anderson metal-insulator transition (MIT)~\cite{Lee85}.
The influence of the disorder on 
the wave functions is reflected by
the mutual correlations between the corresponding energy levels, 
so that the statistics of energy levels  is sensitive to the MIT.
In the metallic limit
 the statistics  of energy spectra 
 can be described 
 by the random-matrix theory (RMT) 
 developed by Wigner  and Dyson~\cite{Dyson62,RMT}.
This was shown 
by solving the zero-mode 
nonlinear $\sigma $-model using the supersymmetric formalism~\cite{Efetov83}. 
Later, perturbative 
corrections 
to the two-level correlation function obtained in the RMT 
were evaluated 
in the diffusive regime 
by the impurity diagram technique~\cite{Altshuler86}. 
In the insulating regime, when the degree of disorder $W$ is much larger 
than the critical value $W_{c}$, 
the energy levels of the strongly localized eigenstates 
fluctuate 
as 
independent
random 
variables.

An important quantity
for analyzing the spectral fluctuations
is the nearest-neighbor level spacing distribution $P(s)$.
It contains information about all of the $n$$-$level correlations.
In the metallic regime
 $P(s)$ 
 is 
very close to the Wigner surmise
$P_{W}(s)=\pi\,s/2\,\exp \left(-\pi\,s^2/4 \right)$~\cite{Imry87}
($s$ is measured in units of the mean level spacing~$\Delta$).
In the localized regime 
the spacings are distributed according to the Poisson law,
{$P_{P}(s)=\exp (- s)$},
because the levels are completely uncorrelated.
The study of the crossover of $P(s)$ between the Wigner 
and the Poissonian limits 
which accompanies the disorder-induced MIT in three-dimensional system (3D)
was started in Ref.~\cite{Altshuler88} and 
became the subject of several subsequent 
investigations
\cite{Shklovskii93,Hofstetter94,Evangelou94,Evangelou92,ZharekeshevK94}. 

It was suggested earlier~\cite{Shklovskii93} that 
$P(s)$ exhibits critical behavior
and should be size-independent 
at the MIT.
Investigating the finite-size scaling properties of $P(s)$ 
provides not only an alternative
method for locating 
the transition~\cite{Hofstetter94}, 
but allows also to determine the critical behavior of the 
correlation length~\cite{ZharekeshevK94}.
A technical advantage of the method 
is that one needs  to compute
only  energy spectra and not 
eigenfunctions and/or the conductivity.
On the other hand, a large number of realizations of the random potential
has to be considered. In comparison with the well-established 
transfer-matrix method~\cite{KramerM94} by which one approaches the MIT
from the localized side, the level-statistics procedure
starts from the metallic regime. 
Thus, the two methods can be considered to provide
complementary information about the critical region.

The suggestion of the existence 
of a third {\em universal} level statistics at the MIT
excited considerable 
interest in the explicit form 
of the critical spacing distribution.
From general considerations for the orthogonal symmetry~\cite{RMT}
$P(s)$$\propto$$s$ at small~$s$.
For large $s$, essentially two
different analytical expressions 
were proposed~\cite{Brody73}.
One of them~\cite{Shklovskii93,Altshuler88} assumes
that 
$P_{c}(s)$ 
is a Poissonian 
for $s\gg 1$,
since
at the critical point 
the Thouless energy, which is a measure of the number of energy levels
 that contribute 
to the average conductance of the system, is of order of 
$\Delta$,
while level repulsion
is important only for small $s$. 

A different asymptotic form, 
$P_{c}(s) \propto \exp(-A\,s^{\alpha})$, 
was proposed~\cite{Aronov94}, 
by using an analogy between the sequence of energy levels and
a classical one-dimensional gas of 
interacting fictitious particles.
Here 
$\alpha$ is
given by
the dimensionality $d$ and
the localization length exponent $\nu$,
\begin{equation}
\alpha=1+(d\nu)^{-1}.
\label{alpha}
\end{equation}
The result is obtained in
the Gibbs model by assuming 
the power law  $s^{2-\alpha}$ 
for the pairwise interaction between the particles~\cite{Kravtsov94}.
The latter distribution
decays faster than the Poissonian  
($\alpha = 1$), 
but slower than the Wigner surmise ($\alpha = 2$). 
Several numerical calculations for the 3D Anderson model
were recently performed~\cite{Evangelou94,HofstetterV94}
in order 
to analyze 
$P_{c}(s)$.
The results were found to be consistent with
the latter of the above suggestions
with an exponent $\alpha \approx 1.2-1.3$ ($\nu\approx 1.5$).
However, since the rounding errors in the calculations for large $s$
are such that
$\alpha=1$ cannot be completely ruled out,
the asymptotic form of $P_{c}(s)$
is still an open question,
and the subject of presently on going and controversial discussions.
In this Letter 
we present the results of  detailed high-precision numerical
investigations of the critical level spacing distribution.
Our findings solve the above controversy.
Preliminary results have been published previously~\cite{ZharekeshevKJpn}.

By diagonalizing the Anderson Hamiltonian  with a 
Lanczos algorithm~\cite{ZharekeshevK94} 
specifically modified  
for  systems
containing up to 10$^6$
lattice sites, 
which were not achieved in previous works, 
we examined both the critical behavior
and the finite-size scaling  properties of the integrated probability
distribution of neighboring spacings $I(s)$.
Our main result is that
the asymptotic form of critical $I_{c}(s)$ and, therefore, $P_{c}(s)$
at large~$s$ is very close to a Poissonian decay, 
as the leading term, 
thus confirming the ideas  of~\cite{Shklovskii93,Altshuler88}. 
In addition, by using the size-independence  of $I_{c}(s)$  at the MIT
and investigating the scaling of $I(s)$ with 
the system size $L$ and $W$,
we estimate the correlation length
exponent $\nu$.
%
%
\begin{table}[t]
\caption[]{Numerical parameters for various cube sizes $L$ at $W_{c}$=16.4. 
$M$: number of samples, $N_{s}$: total number of spacings,
$\Delta$: mean level spacing,
$\rho=(\Delta L^{3})^{-1}$: density of states;
$\alpha$ and $A_{c}$: 
quantities of Eq.~(\ref{Critic}).
All levels lie within the energy interval 
$|E|<4.45$.
}
\label{ttable1}
\begin{tabular}{llrllcc}
$L$ & $M$ & $N_{s}$ & $\Delta$ & $\rho$ &   $A_{c}$ & $\alpha$\\
\hline
 5 & $3\,10^5$ & 18\,610\,321 & 1.42\,10$^{-1}$  & 5.62\,10$^{-2}$ & 1.90 & 1.01 (0.02)\\
 8 & $4\,10^3$ &  1\,016\,790 & 3.47  & 5.63 & 1.89 & 0.95 (0.06)\\
12 & $3\,10^3$ &  2\,576\,306 & 1.03  & 5.62 & 1.89 & 0.99 (0.05)\\
16 &  $5\,10^2$ & 1\,017\,902 & 4.34\,10$^{-3}$ & 5.62 & 1.88 & 0.98 (0.06)\\ 
20 &    25 &      99\,493 & 2.23 & 5.61 & 1.91 & 1.00 (0.10)\\ 
28 &    10 &     109\,075 & 8.11\,10$^{-4}$ & 5.62 & 1.87 & 1.07 (0.10)\\
32 &   10 & 163\,097 & 5.40 & 5.62 & 1.89 & 0.99 (0.08)\\
40 &   5 & 158\,658 & 2.77 & 5.62 & 1.91 & 0.97 (0.09)\\
64 &   2 & 260\,020 & 6.79\,10$^{-5}$ & 5.62 & 1.88 & 1.04 (0.09)\\
80 &   1 & 254\,321 & 3.47 & 5.62 & 1.92 & 1.02 (0.06)\\
100\tablenotemark[1] &   1 & 99\,360 & 1.77 & 5.63 & 1.88 & 0.95 (0.11)\\
\end{tabular}
\tablenotetext[1]{energy interval is $|E|<0.89$.}

\end{table}

The Anderson model~\cite{Anderson58} is defined by
$H=\sum_{n}\varepsilon_{n}^{} a_{n}^{\dag} a_{n}^{} + \sum_{n\neq m}
(a_{n}^{\dag} a_{m}^{} + c.c.)$, where $a_{n}^{\dag}$ ($a_{n}^{}$) is
the creation (annihilation)  operator of an electron at a site $n$,
with  $m$  denoting the nearest neighbors of $n$. The site energies
$\varepsilon_{n}$  are measured in units of the overlap integral
between adjacent sites. They are independent random variables that are
distributed around $\varepsilon=0$  according to a box distribution of
width $W$. 
A simple cubic lattice with periodic boundary conditions was used. 
We computed the electron spectra of cubes 
of linear size ranging from $L$ = 5 to 100 for various $W$. 
It is known from the transfer-matrix
method \cite{Kramer83}, that in the center of the band
($\varepsilon=0$) $W_{c}\approx 16.4$. 
The spectrum was properly ``unfolded'' by
fitting the  integrated density of states around $\varepsilon=0$ to
polynomial splines. The numerical results at the MIT are summarized in 
Table~\ref{ttable1}. 
It should be noted that the numerical
diagonalization of giant sparse matrices of order of $10^5$-$10^6$
is highly nontrivial.

\begin{figure}[tb]
\unitlength1cm
\begin{minipage}[t]{8.5cm}
\begin{picture}(8.5,6.9)
\put(-6.8,-6.4){\includegraphics{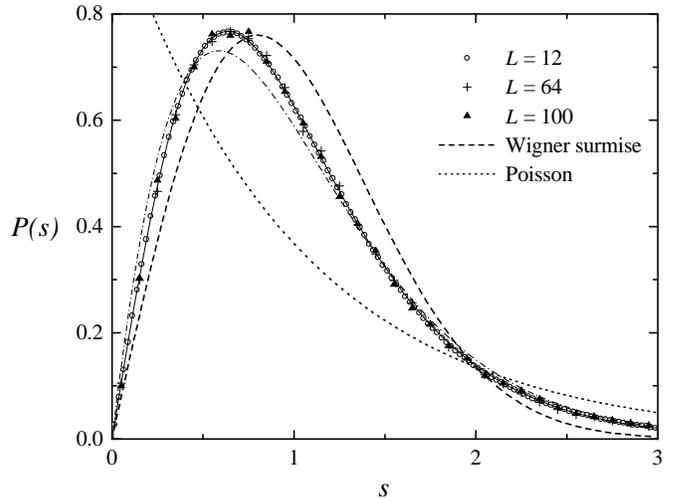}}
\end{picture}
\end{minipage}
\caption[]{Level spacing distribution $P(s)$ 
for various system sizes at the critical disorder $W_{c}$.
Dashed-dotted line is $P_{AKL}(s)$. 
Full line is derivative of $I_{c}(s)$ from 
interpolation formula~(\ref{interpol}).
}
\label{fig1}
\end{figure}

Fig.~\ref{fig1}\, shows $P(s)$ calculated at the MIT. 
As expected, it is $L$-independent. 
%
%
%
To cover the whole range of spacings, the interpolation formula 
$P_{AKL}(s)=B\,s\,\exp(-A\,s^\alpha )$
has been proposed in~\cite{Aronov94}.
Due to normalization
$A=[\Gamma (3/\alpha )/\Gamma (2/\alpha )]^\alpha $ and
$B=\alpha A^{2/\alpha }/\Gamma(2/\alpha )$.
The best fit using the $\chi^2$-criterion 
in the interval $0<s<4$ yields $\alpha = 1.48 \pm 0.08$
with a confidence level 0.95. 
The fitted exponent $\alpha$ is markedly 
larger than that given by~(\ref{alpha}).
For $s>3$
one observes 
an increasing deviation between $P_{AKL}(s)$
and the computed histogram. 
This shows that fitting near
$s \sim 1$ does not provide
reliable information about  $\alpha $, because 
the exponential tail of $P(s)$ contributes 
to the relative accuracy 
only with a very small weight.
Therefore it is imperative
to investigate
the asymptotic behavior at large $s$, not including data from 
the region  $0<s\lesssim 2$.    

In what follows, we consider  the  cumulative level spacing
distribution function 
$I(s)\equiv \int_{s}^{\infty}\,P(s^{\prime})\,ds^{\prime}$.
It gives the probability to find
neighboring energy levels with a separation $E>s\,\Delta$. 
The integration does not change the asymptotic  exponential behavior of $P(s)$. 
Since $s>0$,  $I(0)$=1, and by normalization to the total
number of spacings in a given interval,
$\int_{0}^{\infty}\,I(s)\,ds=1$. 
The Wigner surmise~\cite{surmise}  and the Poisson distribution  yield  
$I_{W}(s)=\exp(-\pi\,s^2/4)$ and
$I_{P}(s)=\exp(-s)$, respectively.  
The numerical evaluation of $I(s)$
is similar to that of the 
density of states in unfolding the spectrum. 
By arranging the spacings in a 
descending sequence one can  very accurately construct the histograms of 
$I(s)$~\cite{Hofstetter94}.

\begin{figure}[tb]
\unitlength1cm
\begin{minipage}[t]{8.5cm}
\begin{picture}(8.5,8.2)
\put(-1.4,-9.0){\includegraphics{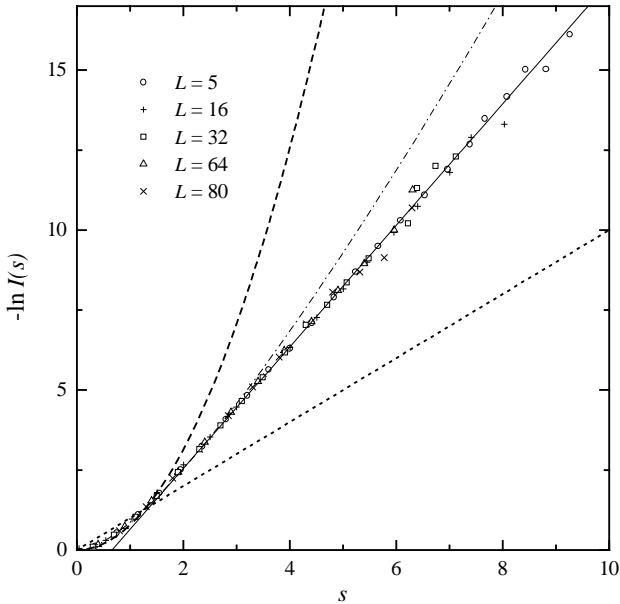}}
\end{picture}
\end{minipage}
\caption[]{Critical probability of neighboring spacings $I(s)$. 
Solid line is Eq.~(\ref{Critic}) with $\alpha = 1$.
Dashed-dotted line is power law with 
$\alpha = 1.24$ from Eq.~(\ref{alpha}).
Dashed and dotted line are $I_{W}(s)$ and $I_{P}(s)$,
respectively.
}
\label{fig2}
\end{figure}

Using the common statistical hypothesis at large $s$
\begin{equation}
\ln I_{c}(s) = - A_{c}\,s^{\alpha }, 
\label{Critic}
\end{equation}
we calculated 
$A_{c}$ and $\alpha $  for various $L$ 
(see Table~\ref{ttable1}).
Independently of $L$ the result is $\alpha =1.0 \pm 0.1$. 
The numerical data of  $\ln I_{c}(s)$ shown in 
Fig.~\ref{fig2}
%
%
%
are better described by a {\em linear law} for $s>3$, so that
$I_{c}(s)\propto \exp(-A_{c} s)$ with $A_{c}=1.9 \pm 0.1$.
This is similar to the insulating regime,
although the decay rate $A_{c}$
is larger than unity due to the level repulsion.
The power law
with the exponent $\alpha \approx 1.2$, 
which was recently obtained~\cite{HofstetterV94} 
by an analysis of the shape of $P(s)$
in the range $0<s<5$ for system sizes  $L \le 21$
deviates from our results for $s\gtrsim 4$.

The linear asymptotic behavior of \ $\ln 
I_{c}(s)$
is in contrast to the power law with $\alpha\approx 1.31$ obtained 
numerically~\cite{Evangelou94} for smaller systems $L \le 12$.
The reason for this discrepancy is the following.
The energy interval $E$ considered in~\cite{Evangelou94} 
is so narrow
that it contains only ten spacings on the average, 
that results in a cut-off of $P_{c}(s)$ 
at $s \approx 10\,\Delta $. 
Thus, some fraction of the spacings $s<10\Delta$ is not taken into account,
causing the faster decay of $P_{c}(s)$.
In our calculations the interval is wide enough,
covering approximately half of all of the eigenvalues.
However, such a choice of 
$E$ does not lead to the undesirable
mixture of the extended and the localized states. This is due to a
peculiarity of the box distribution of the site energies $\varepsilon_{n}$.
It follows from the localization phase diagram 
\{$W_{c},E_{c}$\}~\cite{KramerM94,BulkaK85}, that 
the critical disorder $W_{c}$ 
is almost independent of the energy when $|E_{c}|<6$.
In order to investigate how the width of the energy interval 
influences the level statistics, 
we calculated $I(s)$ for 
$E/\Delta=10^2, 10^3$ and $10^4$, provided that
all levels satisfy the critical condition
$L<\xi(\varepsilon)\propto |\varepsilon/E_{c}-1|^{-\nu}$. 
The results were practically the same within the statistical uncertainties. 
This implies the equivalence of averaging over the spectrum and
over the random potential. 
Indeed, due to diminishing the spacing with the size 
$\Delta \propto L^3$, the averaging for smaller
cubes is performed over many samples, while for our largest systems 
$L$=80 and 100 a single realization without ensemble averaging
is even sufficient to get similar distributions with comparable precision.
For numerically describing a crossover 
between small and large $s$,
we propose an explicit form of 
the new
interpolation function
\begin{equation}
I_{c}(s) = \exp [\mu-\sqrt{\mu^2+(A_{c} s)^2}]
\label{interpol}
\end{equation}
with a coefficient $\mu\approx 2.21$. Although we do not provide a
rigorous analytical prove, it gives the excellent fit 
all over the range of the computed
spacings.
The corresponding
$P(s)$ shown in Fig.~\ref{fig1} fulfills the both normalization
conditions.

\begin{figure}[t]
\unitlength1cm
\begin{minipage}[t]{8.5cm}
\begin{picture}(8.5,5.1)
\put(-0.7,-11.5){\includegraphics{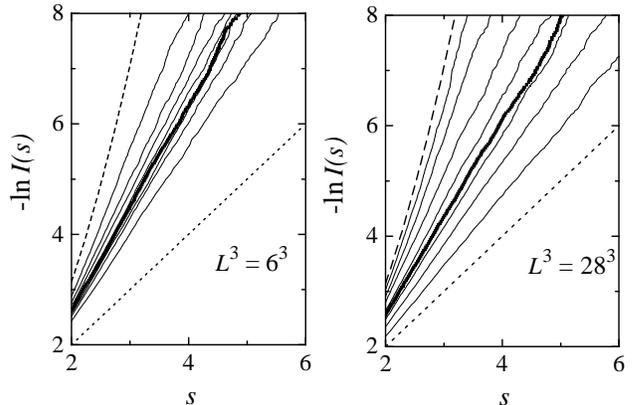}}
\end{picture}
\end{minipage}
\caption[]{Probability $I(s)$ for $L$=6 and 28 at $W$=12, 14, 
15, 16, 16.4, 17, 18 and 20 shown 
consecutively from the left to the right. 
Dashed (dotted) line is the Wigner (Poisson) limit.}
\label{fig3}
\end{figure}

To study the finite-size scaling behavior of $I(s)$ for large $s$,  we
extended the calculations  to other degrees of the disorder $W$  close
to $W_{c}$ for various system sizes. The calculations  were performed
for an ensemble of different samples. The number of samples for each
given pair of $L$ and $W$ was chosen such that  $N_{s} \simeq 10^5$
spacings were obtained.
We have also carefully checked the sensitivity of the results to the
number of realizations. No change was observed within the error bars
when increasing the system size on the expense of the number of
realizations and vice versa.
By increasing $W$  the spacing distribution 
for fixed $L$  changes continuously from $I_{W}(s)$ to $I_{P}(s)$
(Fig.~\ref{fig3}).
%
%
%
The steepness of the crossover 
depends  on $L$.
For larger sizes $I(s)$ changes faster between the 
two limiting regimes.
At $W_c \simeq 16.4$ the spacing distribution 
has almost the same asymptotic form
for all $L$ from 5 to 100.      
This reflects 
the universality of the level statistics 
exactly at the MIT~\cite{Shklovskii93}. 

\begin{figure}[tb]
\unitlength1cm
\begin{minipage}[t]{8.5cm}
\begin{picture}(8.5,7.9)
\put(0.2,-9.7){\includegraphics{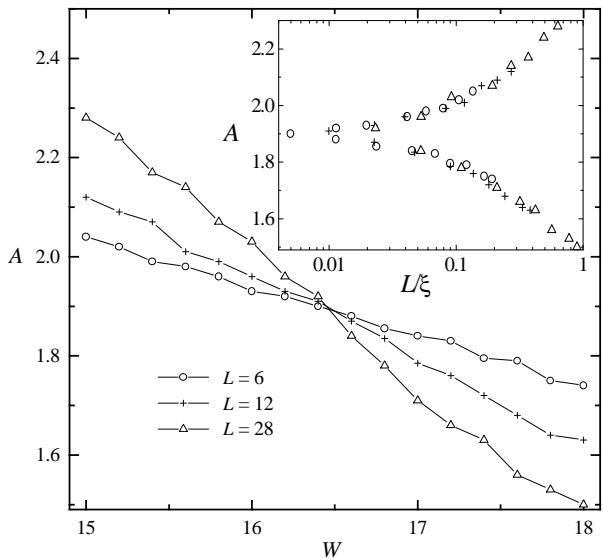}}
\end{picture}
\end{minipage}
\caption[]{Scaling variable $A$ as a function of the disorder $W$
for different $L$, showing critical behavior near the MIT.
Inset: the one-parameter dependence of $A$ on $L/\xi(W)$.}
\label{fig4}
\end{figure}

For finite $L$
the distribution $I(s)$ \ exhibits scaling 
in the vicinity of $W_{c}$.
Within the critical region, $L<\xi(W)$, 
it is reasonable to assume that
the linear slope of \ $\ln I(s)$ \ is governed by 
the one-parameter scaling law, $A(W,L)=f(L/\xi(W))$.
Fig.~\ref{fig4} shows the disorder dependence of $A$ 
near the critical point for various $L$.
%
%
%
$A(W_{c})$ 
does not depend on $L$.
By introducing a scaling parameter, 
the correlation length $\xi (W)$ 
\cite{ZharekeshevK94,KramerM94}, we found 
a common scaling curve consisting of two
branches corresponding to the delocalized and the localized regimes
for $A>A_{c}$ and $A<A_{c}$, respectively, as shown in
the inset of Fig.~\ref{fig4}.
The critical exponent $\nu$ 
was determined in a similar way
as previously~\cite{ZharekeshevK94}, 
where only the small-$s$ part of 
$P(s)$ was used.
We found $\nu=1.4\pm 0.15$ 
in agreement with 
the result obtained earlier by completely different
methods~\cite{Hofstetter94,KramerM94,Kramer83}. 

In conclusion, we present the first large-scale numerical results
on the statistics of the energy levels
near the disorder-induced MIT for systems
of sizes up to \ $L^3=100^3$ sites.
A comparative analysis with results obtained
from various analytical approaches
and other numerical studies is performed.
At the critical point the asymptotic universal
probability of energy level spacings
has a Poisson-like form $I_{c}(s)\propto\exp(-A_{c} s)$.
We believe that the simple exponential asymptotics 
of the critical level spacing distributions
are valid not only for the orthogonal symmetry (with spinless electrons
and without magnetic field), but also for other universality classes:
the unitary (in the presence of the magnetic field) and the symplectic
(in the presence of spin-orbit coupling) classes.
Recent computer simulations~\cite{SchweitzerZ95} 
corroborate that the decay rate 
$A_{c}$ is almost insensitive to the fundamental symmetry.
However, 
it could depend on 
the physical dimensionality.
Finally, we have determined the influence of the 
disorder of the system 
on the exponential tail of $I(s)$, and 
constructed numerically the corresponding scaling function.
The critical exponent of the correlation length 
was calculated, $\nu \approx 1.4$.  

We thank  B.~I. Shklovskii and L.~Schweitzer for discussions.   
This work was supported by the Deutsche Forschungsgemeinschaft
within Project Kr 627, and by the
EU within the SCIENCE, the HCM and the TMR programmes.


\begin{thebibliography}{22}
\bibitem[*]{byline}
E-mail:  {\tt isa@physnet.uni-hamburg.de}

\bibitem{Shklovskii93}
B.~I. Shklovskii, 
{B.~Shapiro}, B.~R. Sears, P. Lambrianides, and H.~B. Shore,
Phys.\ Rev.\ B {\bf 47},  11487  (1993).

\bibitem{Hofstetter94}
E. Hofstetter and M. Schreiber, Phys.\ Rev.\ B {\bf 48},  16979  (1993); {\bf
49}, 14726 (1994).

\bibitem{Kravtsov94}
V.~E. Kravtsov, 
I.~V. Lerner, B.~L. Altshuler, and A.~G. Aronov, Phys.\ Rev.\
  Lett.\ {\bf 72},  888  (1994).

\bibitem{Aronov94}
A.~G. Aronov, V.~E. Kravtsov, and I.~V. Lerner, Pis'ma v ZhETF {\bf 59},  39
  (1994) [{JETP Lett.\ } {\bf 59}, 40 (1994)];                              
Phys.\ Rev.\ Lett.\ {\bf 74}, 1174 (1995);
V.~E. Kravtsov, and I.~V. Lerner, J.\ Phys.\ A: Math.\ Gen.\ {\bf 28},
3623 (1995).

\bibitem{Evangelou94}
S.~N. Evangelou, Phys.\ Rev.\ B {\bf 49},  16805  (1994).

\bibitem{Lee85}
For a review, see P.~A. Lee and T.~V. Ramakrishnan, Rev.\ Mod.\ Phys.\ 
{\bf 57},  287  (1985) and reference therein.

\bibitem{Dyson62}
E.~P. Wigner, Ann.\ Math.\ {\bf 62}, 548 (1955); {\bf 65}, 203 (1957);
F.~J. Dyson, J.\ Math.\ Phys.\ {\bf 3},  140  (1962); {\bf 3}, 1199 (1962).

\bibitem{RMT}
M.~L. Mehta, {\em Random Matrices} (Academic Press, Boston, 1991).

\bibitem{Efetov83}
K.~B. Efetov, Adv.\ Phys.\ {\bf 32},  53  (1983).

\bibitem{Altshuler86}
B.~L. Altshuler and B.~I. Shklovskii, Zh.\ Eksp.\ Teor.\ Fiz.\ {\bf 91},  220
  (1986); [Sov.\ Phys.\ JETP {\bf 64}, 127 (1986)]. 
  Non-perturbative corrections of the two-level correlation function 
  to the RMT for small energies
were found by the combination of the nonlinear $\sigma$-models 
  with the renormalization-group transformation:
  V.~E. Kravtsov and A.~D. Mirlin, Pis'ma v ZhETF {\bf 60},  656
  (1994). 

\bibitem{Imry87}
U. Sivan and Y. Imry, Phys.\ Rev.\ B {\bf 35},  6074  (1987).

\bibitem{Altshuler88}
B.~L. Altshuler, 
I.~Kh. Zharekeshev, S.~A. Kotochigova, and B.~I. Shklovskii,
  Zh.\ Eksp.\ Teor.\ Fiz.\ {\bf 94},  343  (1988);
[Sov.\ Phys.\ JETP {\bf 67},  625 (1988)].

\bibitem{Evangelou92}
S.~N. Evangelou and E.~N. Economou, Phys.\ Rev.\ Lett.\ {\bf 68},  136  (1992).

\bibitem{ZharekeshevK94}
I.~Kh. Zharekeshev and B. Kramer, Phys.\ Rev.\ B {\bf 51}, 17239 (1995-I).

\bibitem{KramerM94}
For a review, 
see B. Kramer and A. MacKinnon, Rep. Prog. Phys. {\bf 56},  1469  (1994) and
references therein.


\bibitem{Brody73}
We do not consider here other interpolation formulae:
T.~A. Brody, Nuvo Cimento Lett.\ {\bf 7},  482  (1973) and 
F.~M. Izrailev,  in {\em Quantum Chaos - Measurement}, Ed. by P.
  Cvitanovic, I. Percival, and A. Wirzba, Kluwer
Scientific Publishers, 1992, p.\ 89.

\bibitem{HofstetterV94}
E. Hofstetter, M. Schreiber, Phys.\ Rev.\ Lett.\ {\bf 73},  3137  (1994);
I. Varga, 
E. Hofstetter, M. Schreiber, and J. Pipek, 
Phys.\ Rev. B {\bf 52}, 7783 (1995).

\bibitem{ZharekeshevKJpn}
I.~Kh. Zharekeshev and B. Kramer, Jpn.\ J.\ Appl.\ Phys.\ {\bf 34} (8B),
4361 (1995).


\bibitem{Anderson58}
P.~W. Anderson, Phys. Rev.\ {\bf 109},  1492  (1958).

\bibitem{Kramer83}
A. {Mac\,Kinnon} and B. Kramer, Z.\ Phys.\ B {\bf 53},  1  (1983).

\bibitem{surmise}
A more exact asymptotic form of $P(s)$ for the GOE
 is {$\ln P(s)=-(\pi\,s/4)^2 - \pi\,s/4 + O(\ln s)$~[8]}.    


\bibitem{BulkaK85}
B. Bulka, B. Kramer and A. {Mac\,Kinnon}, Z.\ Phys.\ B {\bf 60}, 13 (1985);
B. Bulka, M. Schreiber and B. Kramer, Z.\ Phys.\ B {\bf 66}, 21 (1987).
 
\bibitem{SchweitzerZ95}
M. Batsch, {L. Schweitzer, I.~Kh. Zharekeshev} and B. Kramer,
Phys.\ Rev.\  Lett.\ {\bf 77}, 1552 (1996);
T.~Kawarabayashi, 
T. Ohtsuki, K. Slevin, and Y. Ono, Phys.\ Rev.\  Lett.\ {\bf 77}, 3593 (1996).

\end{thebibliography}
\end{document}